\documentclass[]{spie} 
\usepackage[T1]{fontenc}
\usepackage{lmodern}
\usepackage{textcomp}
\usepackage{microtype}
\usepackage{dcolumn} 

\usepackage{amsmath}
\usepackage{natbib}
\usepackage{amssymb}

\usepackage{eqlist} 
\usepackage[final]{graphicx}
\usepackage[dvipsnames]{color}
\DeclareGraphicsExtensions{.pdf, .jpg}

\usepackage{cite}
\usepackage{natbib}
\usepackage{titlesec}
\usepackage{upgreek}
\usepackage{epstopdf}
\usepackage{rotating}
\usepackage{wasysym}

\usepackage{hyperref}
\usepackage{color}
\newcommand{\ms}{m s$^{-1}$ }

\newcommand\arcsec{\mbox{$^{\prime\prime}$}}

\newcommand{\fp}{Fabry-P{\'e}rot}

\newcommand\blfootnote[1]{%
	\begingroup
	\renewcommand\thefootnote{}\footnote{#1}%
	\addtocounter{footnote}{-1}%
	\endgroup
}

\title{A comprehensive radial velocity error budget for next generation Doppler spectrometers}

\author{Samuel Halverson\supit{a,b}, Ryan Terrien\supit{a,b,c}, Suvrath Mahadevan\supit{a,b}, Arpita Roy\supit{a,b}, Chad Bender\supit{a,b}, Gu\dh mundur K\'ari Stef\'ansson\supit{a,b}, Andrew Monson\supit{a,b}, Eric Levi\supit{a}, Fred Hearty\supit{a,b}, Cullen Blake\supit{d,g}, Michael McElwain\supit{e,g}, Christian Schwab\supit{f}, Lawrence Ramsey\supit{a,b}, Jason Wright\supit{a,b}, Sharon Wang\supit{a,b}, Qian Gong\supit{e}, Paul Robertson\supit{a,b,h}
\blfootnote{Certain commercial equipment, instruments, or materials are identified in this paper in order to specify the experimental procedure adequately. Such identification is not intended to imply recommendation or endorsement by the National Institute of Standards and Technology, nor is it intended to imply that the materials or equipment identified are necessarily the best available for the purpose.}
\skiplinehalf
\supit{a}Department of Astronomy \& Astrophysics, The Pennsylvania State University, 525 Davey Laboratory, University Park, USA, 16802; \\
\supit{b}Center for Exoplanets \& Habitable Worlds, The Pennsylvania State University, University Park, PA, USA, 16802; \\
\supit{c}National Institute of Standards and Technology, 325 Broadway, Boulder, CO 80305, USA; \\
\supit{d}Department of Physics \& Astronomy, University of Pennsylvania, 209 South 33rd Street, Philadelphia, PA, USA, 19104; \\
\supit{e}NASA Goddard Space Flight Center, 8800 Greenbelt Road, Greenbelt, MD, USA, 20771; \\
\supit{f}Macquarie University, NSW 2109, Australia; \\
\supit{g}NASA Roman Technology Fellow \\
\supit{h}NASA Sagan Fellow
}

\authorinfo{* Contact author: S.P.H, e-mail: shalverson@psu.edu}
 
\begin{document} 
  \maketitle 

\begin{abstract}

We describe a detailed radial velocity error budget for the NASA-NSF Extreme Precision Doppler Spectrometer instrument concept NEID (NN-explore Exoplanet Investigations with Doppler spectroscopy). Such an instrument performance budget is a necessity for both identifying the variety of noise sources currently limiting Doppler measurements, and estimating the achievable performance of next generation exoplanet hunting Doppler spectrometers. For these instruments, no single source of instrumental error is expected to set the overall measurement floor. Rather, the overall instrumental measurement precision is set by the contribution of many individual error sources. We use a combination of numerical simulations, educated estimates based on published materials, extrapolations of physical models, results from laboratory measurements of spectroscopic subsystems, and informed upper limits for a variety of error sources to identify likely sources of systematic error and construct our global instrument performance error budget. While natively focused on the performance of the NEID instrument, this modular performance budget is immediately adaptable to a number of current and future instruments. Such an approach is an important step in charting a path towards improving Doppler measurement precisions to the levels necessary for discovering Earth-like planets.

\end{abstract}


\section{Introduction}
High precision Doppler velocimeters have enabled the discovery of hundreds of exoplanets over the past two decades. While technological improvements in many aspects of precision spectroscopy continue to push the measurement capabilities of Doppler radial velocity (RV) spectrometers, future instruments will require an exquisite understanding of instrumental and astrophysical noise sources to push towards the discovery of terrestrial mass planets. As such, developing a detailed instrument performance budget is critical for both identifying instrument error sources, and estimating achievable Doppler measurement precision. 

As instruments begin to push below 1 {\ms} measurement precision, many individual error contributions from a variety of subsystems begin to set the measurement floor. Identifying and characterizing each of these contributions represents a significant technological challenge, though one that must be overcome for next generation instruments aiming for 10 c{\ms} precision.

Here we describe the instrumental error budget for the NASA Extreme Precision Doppler Spectrometer \lq{}NEID\rq{} (NN-explore Exoplanet Investigations with Doppler spectroscopy). NEID is a highly stabilized spectrometer for the 3.5 m WIYN telescope, and is designed specifically to deliver the measurement precisions necessary for detection of Earth-like planets. The core of NEID is a single-arm, high resolution (R$\sim$100,000) optical (380 -- 930 nm) spectrometer mounted to an aluminum optical bench. NEID borrows heavily from the design of the Habitable-zone Planet Finder instrument (HPF) \citep{Mahadevan:2014a}, sharing a nearly identical vacuum chamber, radiation shield, and environmental control system \citep{Hearty:2014}.

\section{Methodology}

\begin{figure}
\begin{center}
\includegraphics[width=6.7in]{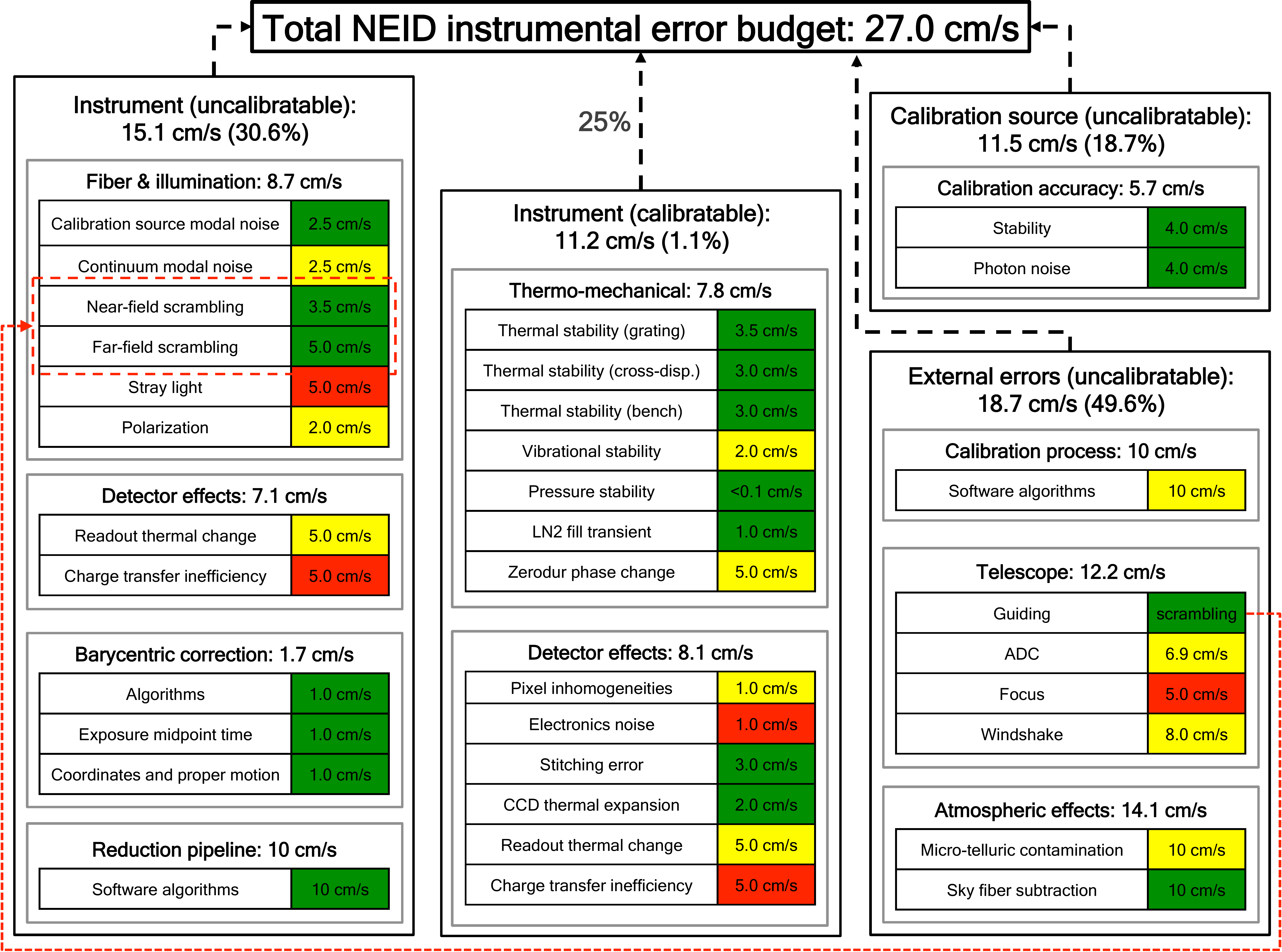}
\caption{Our global NEID error budget table showing individual error contributions from all noise sources described in the text. Major subsystems are broken out in sub-fields with individual contributions to the total error budget listed next to the system labels. Furthermore, all terms are separated using \lq{}calibratable\rq{} and \lq{}uncalibratable\rq{} tags. Calibratable errors are traced by the calibration source, while uncalibratable terms are not. Color codes for each term indicate level of detail in constraining particular error contributions, with well understood (green), moderately constrained (yellow), and more poorly constrained (red) terms listed. The final, single-point measurement precision is derived by combining uncalibratable and calibratable errors in quadrature. We conservatively assume 25\% of the calibratable errors are left uncorrected in the velocity measurement.}
\label{fig:err_budget}
\end{center}
\end{figure}

We base our radial velocity error budget calculations on the thorough and comprehensive systems engineering approach of \cite{Podgorski:2014}, though we consider many additional error sources for NEID such as more specific thermo-mechanical terms (\S~\ref{sec:thermomech}), detailed illumination sensitivity studies (\S~\ref{sec:illumination}), and a more specific characterization of external errors specific to the WIYN telescope (\S~\ref{sec:external}). Figure~\ref{fig:err_budget} shows all individual terms that feed into our final single-point RV measurement error. Error sources are from the spectrometer, wavelength calibration process, telescope and guiding system, and the Earth's atmosphere. Many terms are traced by the dedicated NEID calibration fiber (listed as \lq{}calibratable\rq{}), while others are not (listed as \lq{}uncalibratable\rq{}). We conservatively assume that 25\% of the calibratable errors are left uncorrected, and combine these in quadrature with 100\% of the uncalibratable errors. We show that, even in the event that this percentage is a significant underestimate, the overall budget is generally not dominated by calibratable terms. The different colors used in Figure~\ref{fig:err_budget} trace the degree of constraint on individual error terms. Well characterized or modeled terms are shown in green, while moderately constrained terms are highlighted in yellow. Error sources shaded in red are more poorly constrained, and will require further study to reach a more accurate assessment of actual contribution to the measurement floor.

Some specific noise sources are correlated: these errors do not combine in quadrature and we compensate by holding conservative estimates in these cases. We estimate the final, single-epoch, instrument-limited baseline RV precision error to be $\sim$27 c{\ms} for NEID (excluding astrophysical noise sources). We justify the magnitude of individual terms in the subsections below, and this drives many of our performance requirements on external systems (e.g. telescope guiding precision, spectrometer room environmental control, etc.). Fractional contributions to the total error budget are provided in Figure~\ref{fig:err_budget}, separated in various basis groups. This comprehensive error analysis is essential if NEID is to meet its bold precision goal, and also shows a credible, albeit daunting, path towards 10 c{\ms} through future mitigation and characterization of certain instrumental effects and suppression of external errors. 

The following subsections describe each individual error term in detail and summarize the methods, calculations, or reasoning used to derive individual estimates of error budget contribution.

\subsection{Thermo-mechanical error sources}
\label{sec:thermomech}
Variations in the spectrometer environment, such as temperature variations, pressure changes, and external physical perturbations will all adversely affect the achievable measurement performance. The direct impact of these \lq{}thermo-mechanical\rq{} error sources is difficult to accurately model in the context of Doppler measurement precision, though these sources can be broken into many discrete error terms.

\begin{figure}
\begin{center}
\includegraphics[width=5.2in]{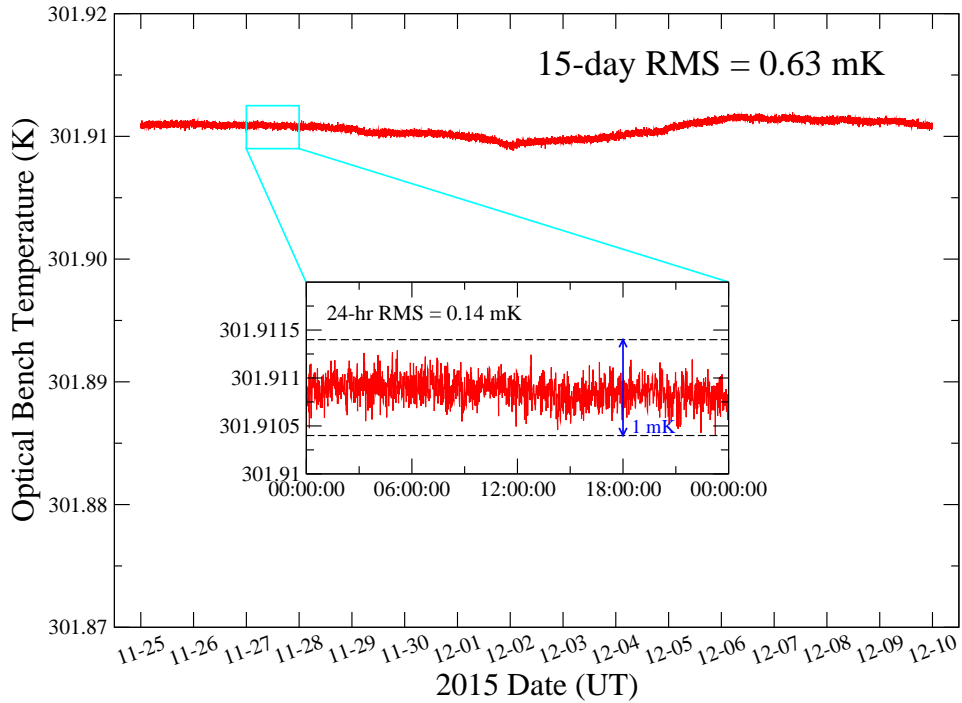}
\caption{Demonstrated thermal control precision of \lq{}warm\rq{} HPF cryostat at NEID operating temperature showing individual control channels (top) and achieved thermal stability on the optical bench (bottom). Sub-mili-Kelvin control is readily achievable in this system, significantly mitigating thermal-induced RV errors. Figure to appear in Robertson et al. 2016 (these proceedings).}
\label{fig:NEID_thermal_demo}
\end{center}
\end{figure}

To estimate the influence of thermal perturbations on RV measurement performance, we adopt a reduction factor of 200 between temperature variations in the spectrometer room (of order $\pm$0.2 K) to expected variations inside the NEID vacuum vessel during standard operation ($<$1 mK). This factor is based on extensive lab testing with the HPF vacuum chamber (\cite{Hearty:2014}, Stef\'ansson et al. 2016 these proceedings), where we routinely achieve a reduction factor of 20 between the outside and inside of the vacuum vessel and an additional factor of 10 after installing a thermal enclosure around the vacuum vessel to further dampen room variations. We have already demonstrated temperature control of the optical bench and radiation shield at the $<$1 mK level with the HPF vacuum chamber, which is nearly identical to the NEID chamber, operating at NEID's temperature setpoint (see Figure~\ref{fig:NEID_thermal_demo}, and Robertson et al. these proceedings), implying these limits are all likely conservative. This factor of 200 is already sufficient for NEID project requirements, but is further improved upon in our design with a fully sealed thermal enclosure and a tighter packing of resistive heaters (compared to the HPF design) to better regulate the temperature of thermal shield surrounding the spectrometer optics. 

\subsubsection{Thermal stability of dispersers} 
\label{subsubsec:thermal_echelle}

Temperature variations change will change the effective length of the echelle grating, even for gratings written on low-expansion substrate materials. This minute length variation will change the effective groove spacing on the active grating surface, leading directly to spectral shifts in the focal plane of the instrument. For a 200 x 800 mm, 31.6 groove/mm R4 echelle grating (a common choice for high resolution Doppler spectrometers) on a pure Zerodur substrate, assuming a highly conservative CTE value for Zerodur of 0.4 ppm K$^{-1}$ at 300 K, a 1 mK temperature change corresponds to a 1.2 c{\ms} equivalent spectral shift in the spectrometer focal plane. We have specifically selected class 0 Zerodur for the NEID echelle substrate material, so we expect the CTE to be significantly lower (of order 0.01 ppm K$^{-1}$)\footnote{CTE values from Schott product page (June 2016): \url{http://www.schott.com/advanced_optics/english/download/schott_zerodur_katalog_july_2011_en.pdf}} for our grating at the 300 K nominal operating temperature of NEID.

In addition to the primary disperser, the transmission properties of the large NEID prism cross-disperser (approximately 360 mm in base length, 260 mm in height) will also change with temperature. Thermal variations will change both the shape and refractive index of the large PBM2Y prism, though the bulk of this change (by definition) shifts the spectrum in the cross-dispersion direction. We simulated the effects of a 1 mK temperature change (single step change) in Zemax and find it contributes $\sim$1 c{\ms} mK$^{-1}$ drift in the dispersion direction. The significantly larger sensitivity in the cross-dispersion direction will contribute minimal additional error to the RVs, assuming a reliable order trace, sufficiently repeatable extraction routines, and a well characterized detector. The thermal effects from both dispersive elements (grating and prism) are traced with the NEID calibration fiber.

\subsubsection{Thermo-elastic stability} 

\begin{figure}
\begin{center}
\includegraphics[width=4.6in]{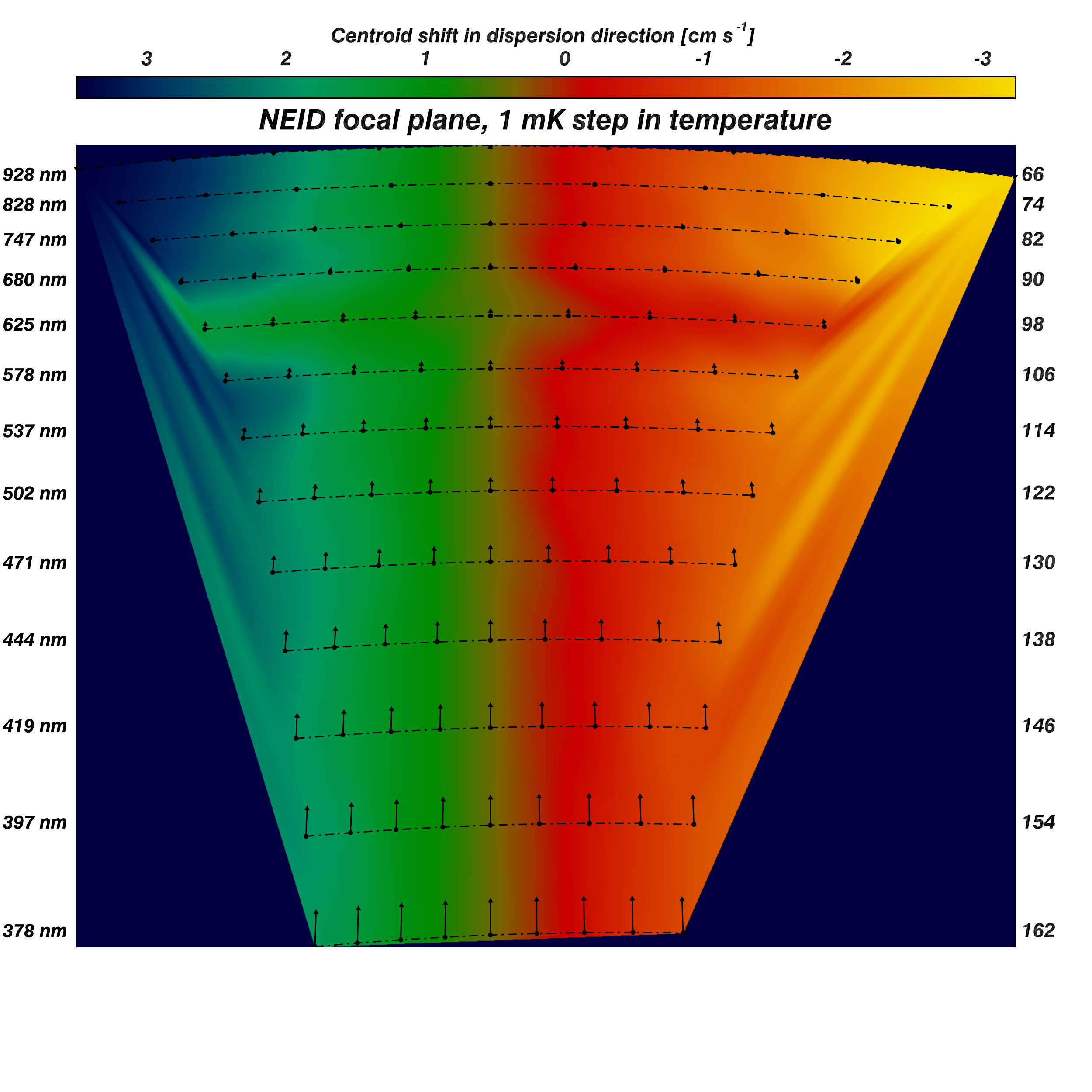}
\vspace{-32pt}
\caption{Simulated NEID focal plane displacements for a 1 mK DC temperature offset for all optical and mechanical components. Arrows trace magnitude and direction of centroid shifts of sampled image points in the focal plane. The excellent image quality and symmetric aberration distribution encoded in the NEID optical design minimizes sensitivity to thermal fluctuations at the mK level.}
\label{fig:NEID_thermal}
\end{center}
\end{figure}

This term accounts for the thermo-elastic nature of the optical bench and optical mounts. As the NEID aluminum optical bench and mounts expand and contract with thermal fluctuations, the recorded spectral features will systematically shift in the focal plane. As a zeroth order test, we expanded the optical bench in Zemax assuming a 1 mK thermal \lq{}step\rq{} and directly measured the effect on individual spots in the NEID focal plane by comparing the ray traces before and after the temperature was increased. This simplified analysis does not take into account detailed optical mounting fixtures, but does provide a reasonable estimate of expected error contribution. Effective centroid shifts of individual spots traced to the focal plane are shown in Figure~\ref{fig:NEID_thermal}, showing minimal sensitivity to DC thermal offsets at the $\sim$1 mK level.

For small changes in temperature, the shifts are approximately linear when averaged over the entire detector (with a 1 mK change corresponding to roughly 1 c{\ms} in the dispersion direction, excluding the thermal sensitivity of the dispersive elements). This assumption of a uniform step in temperature represents a highly simplified model, though is a reasonable approximation when high conductivity materials, such as Aluminum, are used for the optical bench and optic mounts (as is the case with both HPF and NEID). The use of such materials minimizes gradients along the optical bench and mounts, resulting in isothermal heating or cooling during thermal transients.
\begin{figure}
\begin{center}
\includegraphics[width=5.3in]{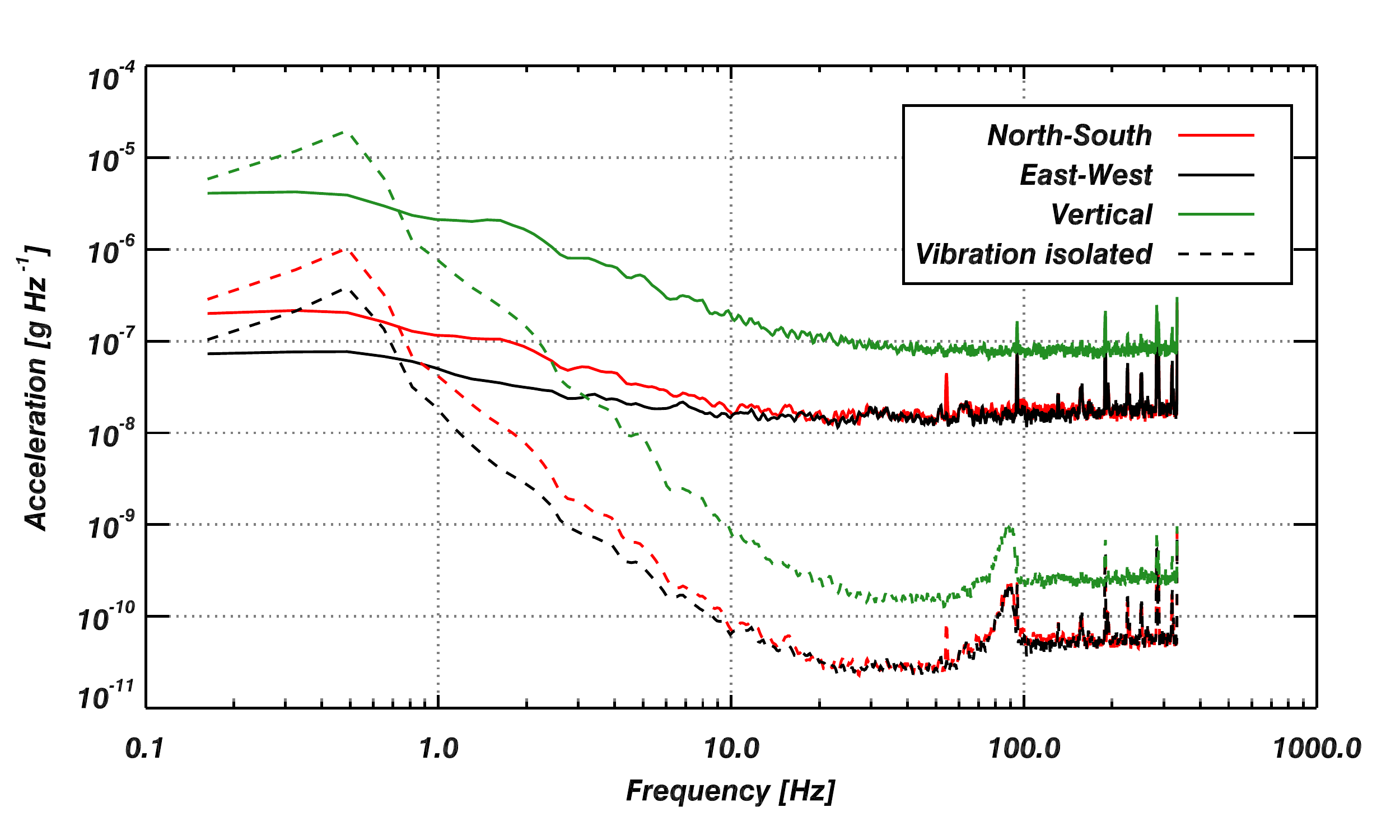}
\caption{Measured North-South, East-West, and vertical acceleration spectrum of WIYN floor in telescope basement (sold lines). Dashed lines are the same spectra modulated by the expected attenuation added by commercial isolation legs. The addition of the active damping legs will improve the stability of the NEID optical bench significantly, particularly at higher frequencies ($>$10 Hz).}
\label{fig:vibration_spec}
\end{center}
\end{figure}

\subsubsection{Vibrational stability}

External vibrations will displace optical components to some extent. Any differential motion between components will directly impact achievable measurement performance. Vibration dampening technologies, such as active isolation legs and passive mechanical decoupling joints, will significantly reduce the net vibrations felt by the instrument optical bench. Translating the expected vibration spectrum felt by the spectrometer to an estimate of performance degradation can be difficult, though we have attempted a simple \lq{}worst-case\rq{} study here.

For NEID, the vibration spectrum felt by the optical bench will be that of the room, modulated by an active vibration damping system. We used the measured vibration spectrum measured of the WIYN instrument room floor\footnote{WIYN spectrometer room acceleration data from (Jan 2016): \url{http://www.wiyn.org/About/wiynEPDS.html}} and applied the suppression spectrum of typical commercial air legs (in this case, we used the dampening spectrum of a commercial Minus-K Newport isolator series) to calculate a total integrated (NS/EW/vertical) acceleration of $\pm$10 $\upmu$g. Figure~\ref{fig:vibration_spec} shows the measured and attenuated vibrational power spectra at the WIYN spectrometer room. Conservatively mapping this entire added acceleration to a corresponding deformation at the center of the optical bench, we measure an effective shift in the spectral direction of $\sim$1 c{\ms}. This simplified model conservatively assumes that the detector assembly moves independently from the rest of the optical train, meaning this is likely an upper-limit for this particular error contribution.

\subsubsection{Vacuum chamber pressure stability}
\begin{figure}
\begin{center}
\includegraphics[width=5.8in]{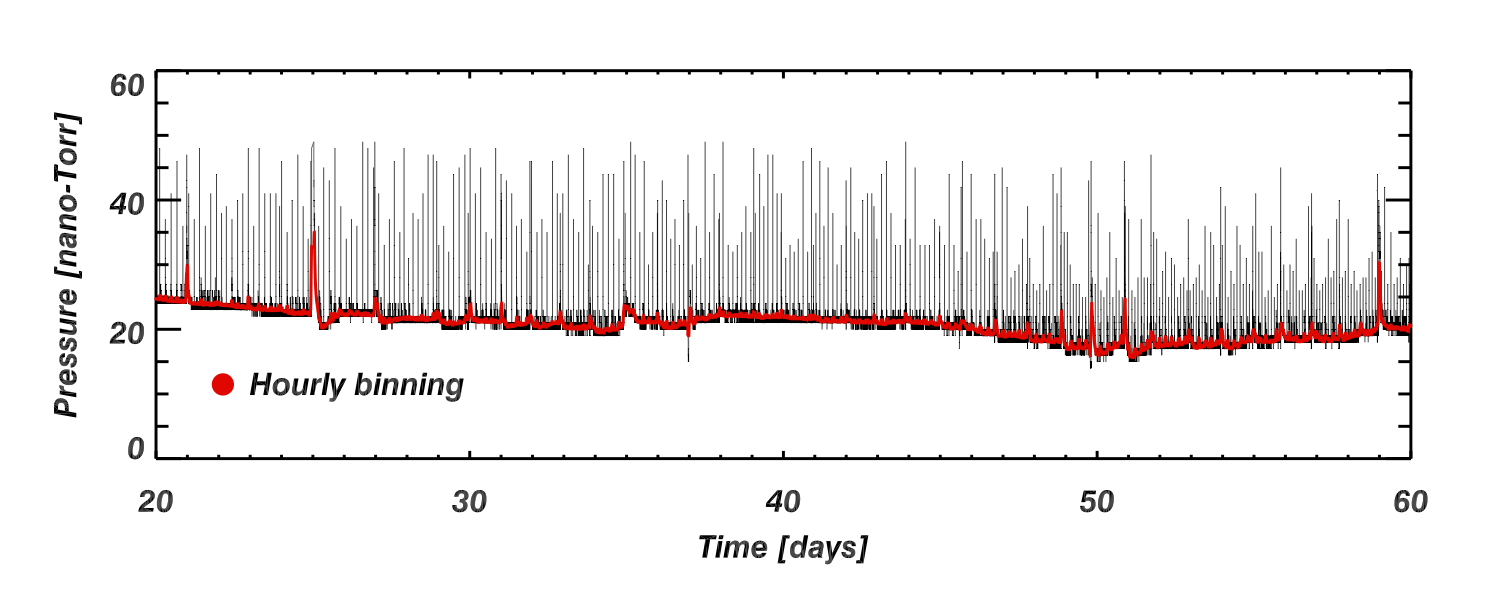}
\caption{Measured long-term pressure stability of HPF/NEID cryostat under full vacuum. The exquisite levels of vacuum stability reduce pressure-related RV errors due to refractive index changes to $<$0.1 c{\ms} under standard operating conditions. The periodic spikes in pressure are likely due to small bursts of outgassing materials within the chamber, though are quickly suppressed by the charcoal getter system.}
\label{fig:pressure_ex}
\end{center}
\end{figure}

Pressure variations within the instrument vacuum chamber will change the effective refractive index of the medium surrounding the spectrometer optics. This directly leads to (calibratable) shifts in wavelengths of recorded spectra. The NEID vacuum chamber design, drawing from significant heritage with the HPF and APOGEE chamber designs, enables an absolute pressure of $<$1 $\upmu$Torr to be maintained over many years of operation. Based on extensive measurements with the APOGEE \citep{Wilson:2012} and HPF cryostats \citep{Hearty:2014}, we expect short-term (24 hour) variations of $<$0.01 $\upmu$Torr under standard operating conditions, translating to a calibratable 0.05 c{\ms} velocity shift due to refractive index changes of the residual medium inside the vacuum chamber. The measured pressure stability within the HPF vacuum chamber is shown in Figure~\ref{fig:pressure_ex}.

The inclusion of a non-evaporable getter and ion pump in the HPF vacuum chamber (nearly identical to the NEID chamber design) maintained a steady vacuum at the $<$0.1 $\upmu$Torr level, effectively eliminating any refractive index changes at measurable levels (see Stef\'ansson et al. these proceedings). We conservatively assign an error of 0.1 c{\ms} to account for both short and long-term (multi-year) pressure gradients over the lifetime of the instrument.

\subsubsection{LN2 fill transient} 

NEID will use liquid nitrogen (LN2) to cool both the detector assembly and the specialized charcoal getters. These getters are essential for maintaining high quality vacuum, as they act as natural absorbers (when cooled to cryogenic temperatures) for residual molecules penetrating the seals used along the chamber walls. The saturation temperature of the LN2 coolant will vary by $\pm$0.2 K in accordance with local fluctuations in the external barometric pressure, introducing a mild thermal transient during each cryogen reservoir filling cycle. During the requisite daily cryogen fills, the introduction of slightly warmer LN2 under pressure will introduce a brief temperature transient into the LN2 tank and detector cold finger.  To minimize this source of variation, NEID will have a permanently installed back-pressure regulator to keep the LN2 in the instrument under a constant pressure (780$\pm$2 Torr), resulting in a relatively constant LN2 saturation temperature of 77.95$\pm$0.02 K. Additionally, cryogen fills will be done during daytime hours, once per day, to maintain temperature of the detectors and vacuum charcoal getters, minimizing impact on science. To account for this error we conservatively assign 1 c{\ms} RV to this term due to the periodic LN2 filling, though further studies will be done to gauge the amplitude of this added thermal transient once the NEID vacuum chamber is fabricated.

\subsubsection{Zerodur phase change} 
Zerodur is made-up of a complex mixture of both glass and ceramic. The precise mixture of glass to ceramic dictates the overall bulk material CTE. Over time, the material undergoes a gradual glass-crystal phase change between the crystalline ceramic and the glass \citep{Bayer-Helms:1987}. This slow phase change manifests as a change in overall length of the grating substrate, which in turn alters the effective diffraction groove spacing over time. The estimated change in length for a typical piece of Zerodur that would be used for an R4 echelle mosaic (200 mm W $\times$ 800 mm L) translates approximately to a calibratable 5 c{\ms} day$^{-1}$ drift (18 {\ms} yr$^{-1}$) velocity offset. This effect has been measured with HARPS (Hans Dekker, priv. comm.), and may be reduced by using aged or specially annealed Zerodur for the grating substrate, rather than newly formed material \citep{Bayer-Helms:1987}.

\subsection{Fiber feed \& illumination}
\label{sec:illumination}

This section outlines error sources related to instrument illumination stability. Delivering a stable illumination to the spectrometer from the telescope is vital for precision Doppler spectroscopy, though achieving stability required for $<$1 {\ms} measurements remains a significant challenge. These errors are generally not traced with the NEID calibration source, and therefore must be minimized independently if the overall precision goal is to be met.

\subsubsection{Modal noise contributions} 
Typical optical fibers used in astronomical applications support a finite number of propagation modes. These propagation modes interfere at the fiber exit boundary, leading to a speckle pattern at the output that depends on the coherence of incident illumination, measurement wavelength, and source bandwidth. This resultant speckle pattern changes as the mode distribution within the fiber is perturbed (e.g. due to telescope movement or temperature variations), and places a fundamental limit on achievable SNR and RV precision \citep{Baudrand:2001, Lemke:2011}. It is important to note that this \lq{}modal noise\rq{} is generally a more prominent issue for wavelength calibration sources, rather than stellar illumination. This is due to the significantly higher coherence of typical high performance frequency references (such as optical frequency combs and broadband etalons) compared to continuum dominated sources (such as starlight, \cite{Mahadevan:2014a, Lemke:2011}). While modal noise can limit achievable SNR on stellar spectra at longer wavelengths \citep{Origlia:2014}, bulk agitation of the NEID science fiber is likely sufficient for minimizing this effect for the stellar illumination.

The HARPS collaboration recently tested a broadband optical frequency comb, developed by Menlo Systems GmbH, by illuminating both the science and calibration channels simultaneously (Menlo Systems, priv. comm.) With the broadband LFC illuminating both HARPS channels, the achieved fiber-to-fiber tracking precision was measured to be $<$2.5 c{\ms} (excluding photon noise, Menlo Systems, priv. comm.). This test relied on heavily agitated fibers to suppress modal noise in both channels, though served as a useful upper limit to modal noise contribution. We expect the NEID calibration fiber, fed by our custom laser frequency comb (also developed by Menlo Systems) though our specialized optical coupling scheme \citep{Halverson:2014b}, to contribute no more than this value to our overall budget. We hold 2.5 c{\ms} RV error for modal noise in both the science and calibration fibers to maintain contingency.

\subsubsection{Near-field scrambling}
\label{sec:nf_scrambling} 
Multi-mode optical fibers are well known to be imperfect image scramblers \citep{Heacox:1986,Avila:2006,Avila:2008}. Any changes in the near-field illumination exiting the NEID fibers will directly manifest as uncalibratable velocity shifts in the spectrometer focal plane. As such, high levels of image \lq{}scrambling\rq{} are required to desensitize the instrument from external illumination variations at the telescope input.

\begin{figure}
\begin{center}
\includegraphics[width=5.2in]{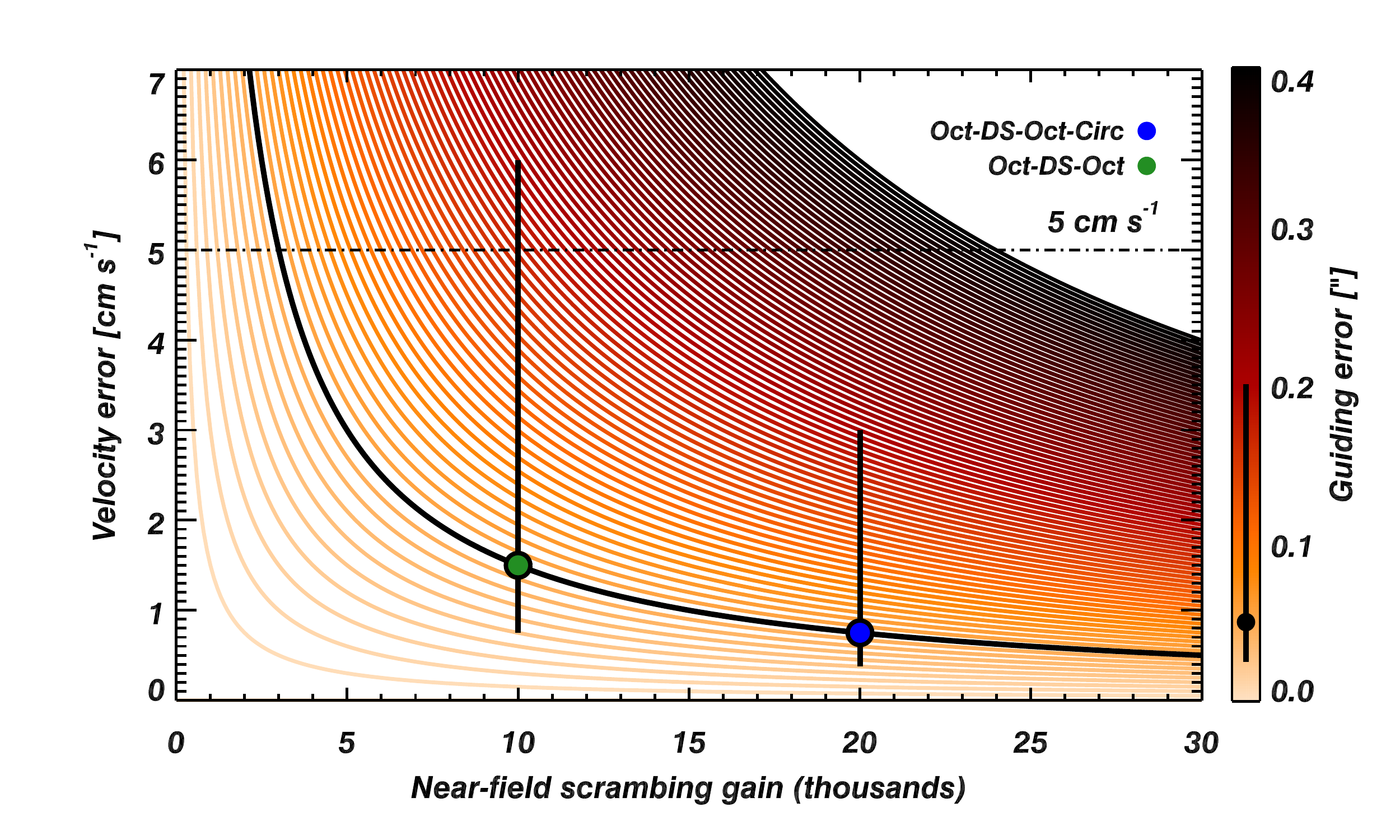}
\caption{Theoretical guiding-induced velocity error for NEID for a range of guiding precisions. Colored curves represent different guiding errors. The solid black line is the expected velocity error based on the NEID front-end guiding precision (0.05{\arcsec} RMS). The colored dots are theoretical velocity errors associated with the near-field scrambling measurements for fiber configurations measured in the laboratory \citep{Halverson:2015a}. Vertical bars show an estimated range of possible guiding precisions for the NEID during a given exposure. For the expected NEID front-end guiding precision (0.05{\arcsec}), the residual velocity error is $<$2 c{\ms}, though we hold a higher value to maintain technical contingency.}
\label{fig:scrambling_nf}
\end{center}
\end{figure}

Our tested HPF fiber delivery system, which includes octagonal fibers, circular fibers, and a specialized double-scrambler system \citep{Halverson:2015a}, yields near-field scrambling gains (ratio of illumination offset at the fiber input input versus the centroid offset at the fiber output) in excess of 10,000 (as measured in the laboratory, \citep{Halverson:2015a}). NEID will use the same fiber configuration as HPF, though use significantly smaller fibers (300 $\upmu$m for HPF vs. 62.5 $\upmu$m for NEID). Changes in the near-field illumination on the fiber input are primarily due to guiding errors and uncorrected chromatic dispersion, so a formal scrambling error is accounted for under both items. Figure~\ref{fig:scrambling_nf} shows the expected near-field scrambling error, calculated using the equations presented in \cite{Halverson:2015a}, for the assumed scrambling gain of the NEID fiber delivery system and expected WIYN guiding precision. Based on measured laboratory scrambling measurements, we expect this term to contribute $<$3.5 c{\ms} error (with some margin) to the overall budget for the expected guiding precision delivered by the NEID front end injection system at the WIYN telescope (0.05{\arcsec}).

\subsubsection{Far-field scrambling} 
Variations in the spectrometer pupil illumination, due to changes in the far-field output of the fiber delivery system, will result in a change in the distribution of aberrations present in the optical system. This variable aberration distribution will lead to systematic PSF shifts in the focal plane that are not traceable by the calibration fiber.

\begin{figure}
\begin{center}
\includegraphics[width=5.3in]{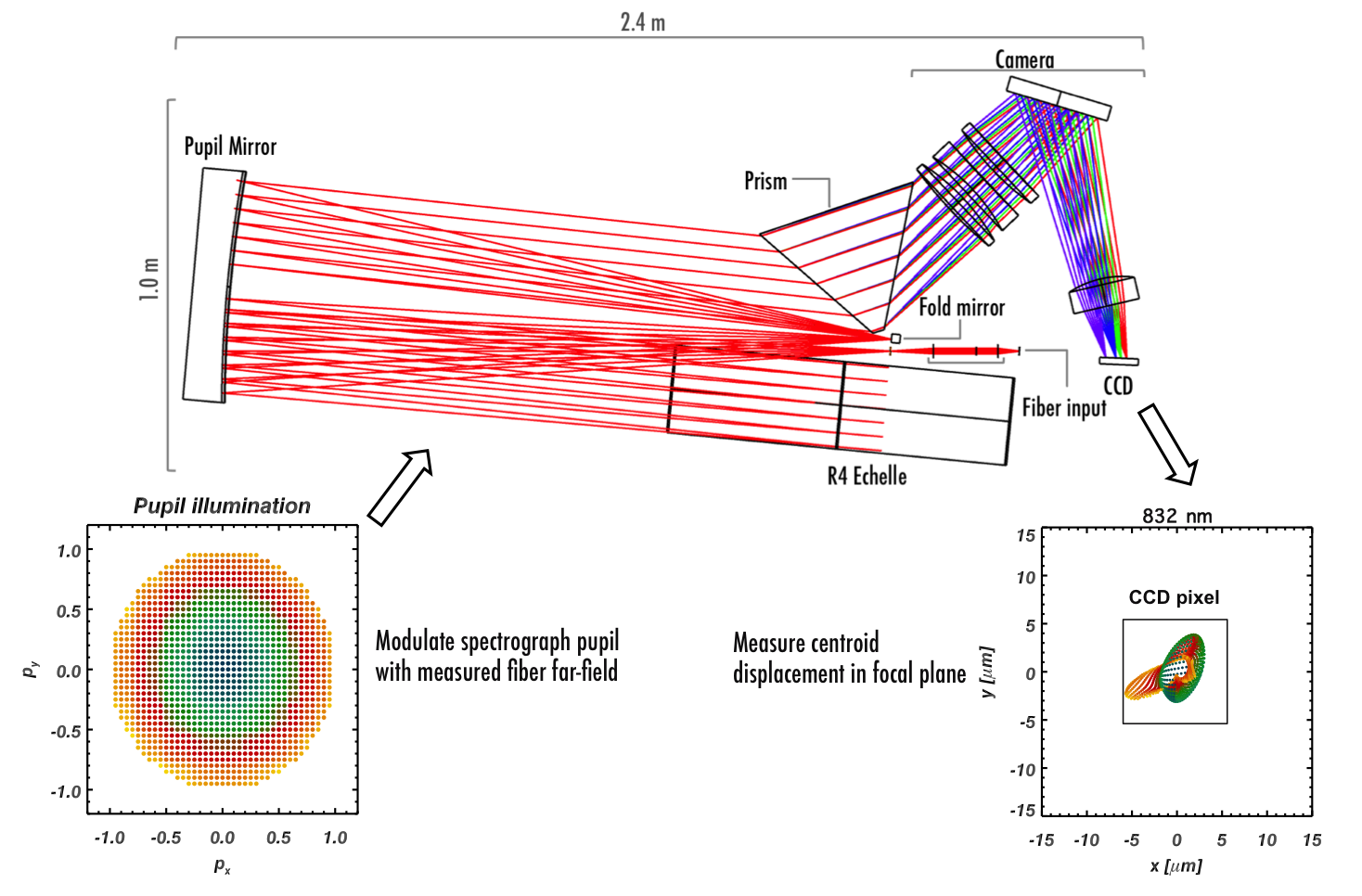}
\includegraphics[width=4.7in]{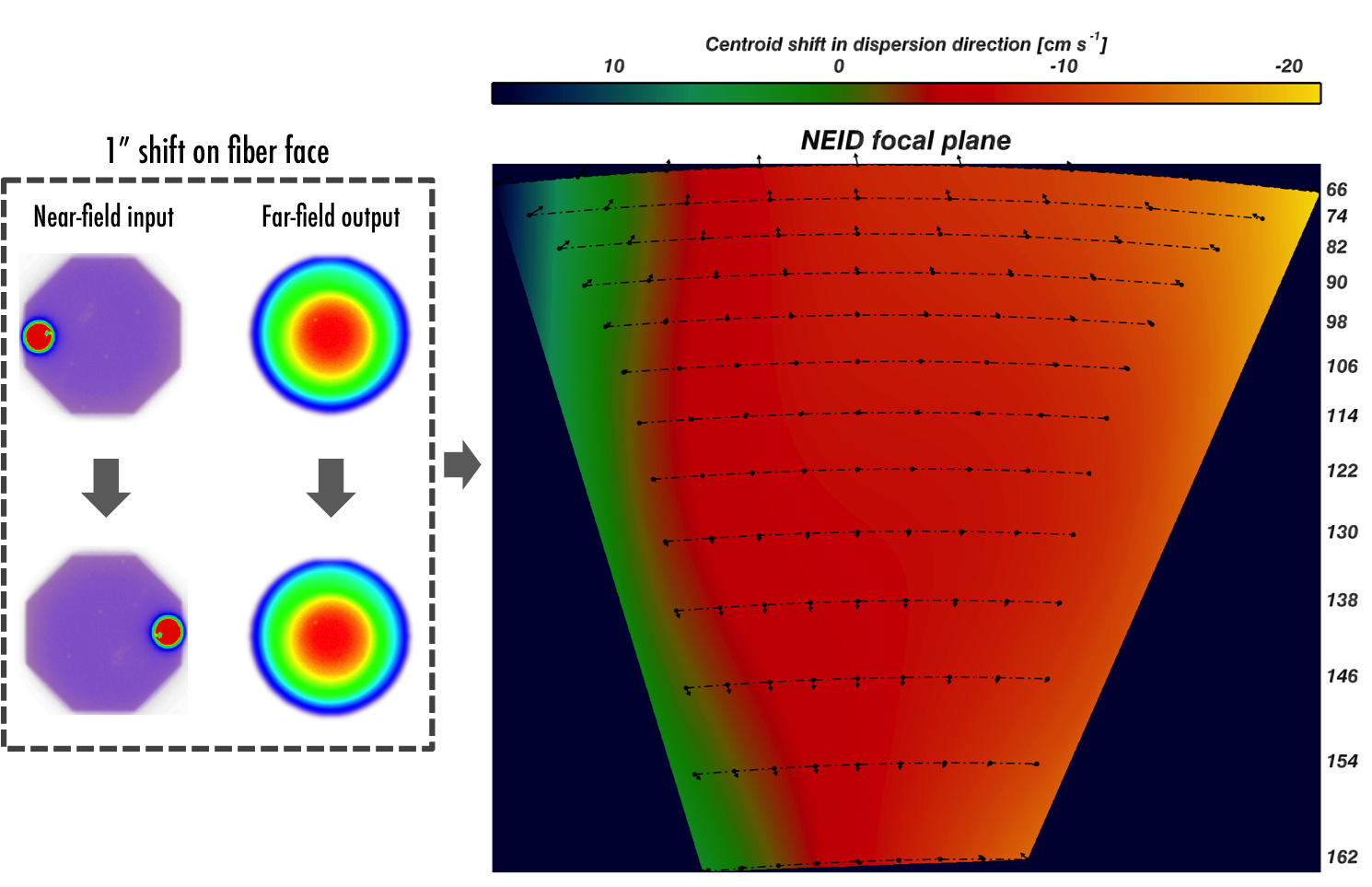}

\caption{Top: Overview of far-field trace method applied to NEID (methodology detailed in \cite{Sturmer:2014}). The measured fiber far-field outputs are used to modulate the spectrometer pupil illumination with a variable weighting map. Weighted rays are then traced to the focal plane for centroid determination. Weighting the instrument pupil differently will result in a change in the effective aberration distribution, leading to centroid shifts in the individual spots in the focal plane. These centroid shifts manifest as spurious Doppler shifts that are not calibratable. Bottom: NEID focal plane showing a subset of spots traced to the detector (echelle orders are listed on the right of the plot). The color scale traces the amount of PSF shift. In this particular example, the output far-fields of an octagonal fiber and double-scrambler system, illuminated at extreme an guiding offset, is used to weight the spectrometer pupil. }
\label{fig:trace_ff_NEID}
\end{center}
\end{figure}

To gauge the expected amplitude of these shifts for the NEID fiber delivery system, we trace the laboratory-measured far-field intensity patterns of the HPF fiber train (\cite{Halverson:2015a}) through the NEID optical model to the detector plane for a variety of different input illumination conditions. Each measured fiber output sets the relative weighting of rays traced in the optical model. Figure~\ref{fig:trace_ff_NEID} shows our overall ray tracing process. This technique is adapted from the methods detailed in \cite{Sturmer:2014}. We then measured the centroid shift of traced spectrometer PSFs across a wide range of wavelengths, each for several different fiber input conditions that emulated different expected input variations. These variations included guiding offsets, telescope pupil variations, and chief-ray angle tilt. Each of these perturbations affects the fiber output in a slightly different way, leading to different far-field intensity patterns. As the NEID fiber system relies on a double-scrambler system for image stabilization, which by definition interweaves the near and far-fields between fibers, it is important to characterize both the near and far-field fiber outputs extensively for different input illuminations.

For a 1{\arcsec} guiding offset (roughly the diameter of the entire NEID fiber), the expected PSF centroid shift due to fiber far-field is $\sim$20 c{\ms} across the CCD. As this extreme guiding error is far above the expected instrument front-end guiding precision, we conservatively hold a small fraction of this number in our overall budget. We have also quantified the sensitivity of this fiber system to incident beam angle variations (i.e., chief-ray angle variations) and incident pupil changes, both which drive the requirements for NEID front end guiding system. All of these contributions are rolled-up into this single error term with $<$5 c{\ms} contribution. This low sensitivity to spectrometer pupil variations is a direct result of both the excellent image quality in our spectrometer design ($>$99\% encircled energy per pixel over the majority of the bandpass) and the high levels of scrambling in our proposed fiber delivery system.

\subsubsection{Stray light}
Our symmetric white-pupil spectrometer design enables efficient suppression of stray light with proper baffling, though we have not explicitly modeled major stray light contributions. A full stray light analysis and baffle design will be part of a future instrument design phase. Given our combination of a tilted detector (over 15$^o$ relative to the back face of the last camera element), large back focal length ($>$100 mm), and no small gaps between lenses, we expect very minor ghosting ($<$3$\times10^{-10}$ fractional irradiance, verified by ghost analysis of all double bounces). We assign 5 c{\ms} error to account for scattered light contamination, though further studies must be done to arrive at more constrained estimates. Proper balancing of calibration and stellar light will be crucial to maintain a constant background light level.

\subsubsection{Polarization}
Variable output polarization can cause systematic shifts in measured radial velocities for spectrographs using echelle reflection gratings as the primary spectral disperser \citep{Halverson:2015b}. Imperfections and internal stresses in the NEID fiber cable will lead to variable coupling between polarization modes (in addition to the spatial modes) within the fiber. This effect is comparable to the more widely studied \lq{}modal-noise\rq{}, though manifests as a change in diffraction efficiency across individual echelle orders, rather than an illumination profile change on the fiber face.  While the NEID fibers fundamentally support thousands of modes (each with multiple polarizations), any low-amplitude residual polarization fluctuations in the fiber will change the effective flux levels across each order. This variable weighting changes both the effective line centroids of spectral features, and the weights of the mask features used to derive the cross correlation functions (CCF). It also influences the overall weighting of a given order CCF, which determines how the individual echelle order CCFs are combined to derive a final velocity measurement. This effect can likely be mitigated with reliable blaze normalization and consistent weighting of spectral features when calculating the CCFs, though we hold 2 c{\ms} error to account for this term until further experiments and simulations are developed.

\subsection{Detector effects}
NEID will use a large format, 9k $\times$ 9k e2v CCD to simultaneously record spectra spanning 380 -- 930 nm. We have identified major sources of error attributed to the NEID CCD detector, though small-scale systematic effects in large-format CCDs are largely uncharacterized at the 10 c{\ms} level. Many of these effects have been identified and studied by the community (e.g. \cite{Bouchy:2009,Molaro:2013}), though further studies of detector-related noise sources will be paramount for next generation instruments aiming for the highest levels of precision. We fully intend to arrive at more accurate estimates of detector-related noise sources in the future with prolonged testing. Nevertheless, we present several terms that have been discussed previously in the literature and include approximate contributions to the overall performance budget for each term. 

\subsubsection{Stitch boundaries}
The photolithographic process for making large format e2v CCDs requires discrete stepping of a precisely constructed lithographic mask. This periodic mask stepping results in measurable positional errors of pixel placement every 512 $\times$ 512 pixels (roughly the size of the base lithographic mask). Pixels on either side of each of the mask boundaries will inevitably have slightly different dimensions, leading to a discontinuity in the otherwise relatively smooth instrument wavelength solution \cite{Molaro:2013, Dumusque:2015}. We assign 2 c{\ms} error based on previous studies with HARPS \citep{Molaro:2013} to account for residual CCD stitching boundary effects, though this effect is readily calibratable with modern frequency comb calibration sources. We plan to characterize the stitch boundaries of the NEID CCD prior to installation in the spectrometer using a combination of interferometric illumination and repeated laboratory tests.

\subsubsection{Thermal fluctuations of silicon absorption layer}
Thermal variations in the detector bulk material (silicon) and mounting interface (silicon carbide) are also a potential source of systematic RV errors. Changing the temperature of the absorbing layer or detector backing will cause physical movements of the pixels across the array. As the temperature of the detector varies during an exposure or readout, the relative size of both the Silicon absorbing substrate and the Silicon carbide backing material changes. This transient expansion of the detector will affect the recorded spectra and lead to degraded RV measurement performance. 

We expect to maintain better than $\sim$10 mK thermal stability of the CCD during typical exposures with commercial temperature controllers. Assuming a typical CTE value for the Silicon absorption layer (roughly 1.2 $\times$ 10${^-6}$ K$^{-1}$ at -100 $^o$C, \cite{Middelmann:2015}) and a 9k $\times$ 9k detector with 10 $\upmu$m pixel pitch, this temperature swing results in an absolute dimension change of $\sim$0.1 nm in the total size of the detector substrate. Converting this dimensional change to an approximate velocity offset using the appropriate dispersion values for NEID (equivalent to roughly 600 {\ms} per 10 $\upmu$m detector pixel) leads to an effective velocity error of {0.5 c{\ms}}. Based on this number, we assign a 2 c{\ms} calibratable error to maintain margin to account for the monolithic, uniform thermal expansion and contraction of the CCD under standard temperature control. 

\subsubsection{Deformations during readout}
Large format CCDs undergo significant thermal fluctuations during readout. The magnitude of these fluctuations depends on the properties of the clocking signals used to read out the array. This added thermal load can result in a measurable warping of the detector, leading to spurious centroid shifts of features in the focal plane. Empirical measurements of non-uniform thermal fluctuations during clocking of large format CCDs \citep{Manescau:2010} suggest detector shape deformations on the order of $\sim$0.15 nm mK$^{-1}$ are present during standard readouts, corresponding to a 2.5 c{\ms} velocity error for NEID with the expected thermal stability of the CCD during readout. We hold twice this value to maintain reserve in our performance budget, though this effect may be mitigated with novel clocking schemes to maintain constant thermal loads. Maintaining a constant thermal load via readout of \lq{}dummy\rq{} output channels will be explored during the NEID detector characterization phase.

\subsubsection{Charge transfer inefficiency}
While fundamentally highly efficient devices, CCDs natively suffer from imperfect charge transfer when shuffling recorded charge between pixels during readout. This charge transfer {\em inefficiency} (CTI) is both spatially variable and flux-dependent. CTI is generally present at the $\sim$0.0010 -- 0.0001\% level (e.g. for a 10,000 ADU signal, a detector CTI of 0.0001\% corresponds to a loss of 10 electrons per 1000 pixel transfers across the array). This effect has long been studied in space-borne instruments where prolonged detector degradation, e.g. due to impurities and surface defects on the CCD caused by cosmic rays, leads to slow deterioration in effective transfer efficiencies \citep{Massey:2010, Massey:2014}.

In the context of precise high resolution stellar spectroscopy, this imperfect transfer of recorded charge introduces minute shifts in the extracted spectral features \citep{Bouchy:2009}. However, CTI effects can be minimized operationally by matching the SNR of different exposures, and by applying flux-defendant empirical corrections during extraction \citep{Bouchy:2009}. We have adopted a loosely constrained error of 5.0 c{\ms} to account for CTI effects, though further study will be done during the NEID detailed design review to attempt to gauge the full magnitude of CTI on the achievable measurement precision for a variety of different expected SNR values.

\subsubsection{Other detector effects} 
Electronics noise and pixel inhomogeneities (not due to the already discussed lithographic stitch boundaries) will inevitably contribute to the final Doppler measurement error, though the direct impact on the RVs is difficult to quantify. Each is included with a loosely constrained 1 c{\ms} estimated contribution in our overall budget. The pixel inhomogeneities can be precisely characterized with our frequency comb calibration source or a precisely positioned interferometric fringe pattern in a similar fashion to the detector stitch boundaries.

To minimize electronics noise, we plan to use the high performance Archon electronics system from STA Inc to interface with our e2v detector. The Archon unit can read out 16 channels through four analog-to-digital conversion boards. We plan to use eight of those channels to read the CCD, and eight to read the dummy channels to maintain constant power during both integration and readout. Our nominal operating mode will use eight channels to read out at 500 kHz. This mode will provide full reads of the array in 20 seconds with read noise of approximately 4 e-, well below the expected noise from other detector-related sources. A slow readout mode will provide 2.5 e- read noise while reading out the full array in 200 seconds.

Light from the dedicated calibration fiber will land on different pixels than the science light, so we have conservatively included the CCD thermal warping and CTI terms (discussed previously) as both calibratable and uncalibratable error terms. Periodically illuminating both the primary NEID science channel and calibration channel with the calibration source will provide relative corrections between the two sets of illuminated pixels to some extent. Constraining these effects will be part of the rigorous NEID detector testing and verification phase.

\subsection{Telescope \& guiding errors}
\label{sec:external}
This section details external errors associated with the WIYN telescope, NEID front-end system, and atmospheric dispersion compensator. Each presents unique additions to the error budget, though we have attempted to quantify the magnitude of each term with realistic estimates.

\subsubsection{Guiding} 
Imperfect guiding will lead to spurious changes in the measured profiles of spectral features. As such, tight guiding requirements and excellent image scrambling performance are paramount for next generation Doppler instruments. A 0.05{\arcsec} RMS guiding precision on the 0.92{\arcsec} NEID fiber translates to 0.9 c{\ms} guiding-induced error, assuming a near-field scrambling gain of 20,000 for the fiber delivery system. We carry four times this value (equivalent to assuming a 0.10{\arcsec} RMS guiding precision with a near-field scrambling gain of 10,000) as technical contingency for periods of degraded guiding or adverse weather. This term is carried under \lq{}scrambling\rq{} in the overall performance budget.

\subsubsection{Atmospheric dispersion}
\label{subsubsec:adc} 
\begin{figure}
\begin{center}
\includegraphics[width=5.2in]{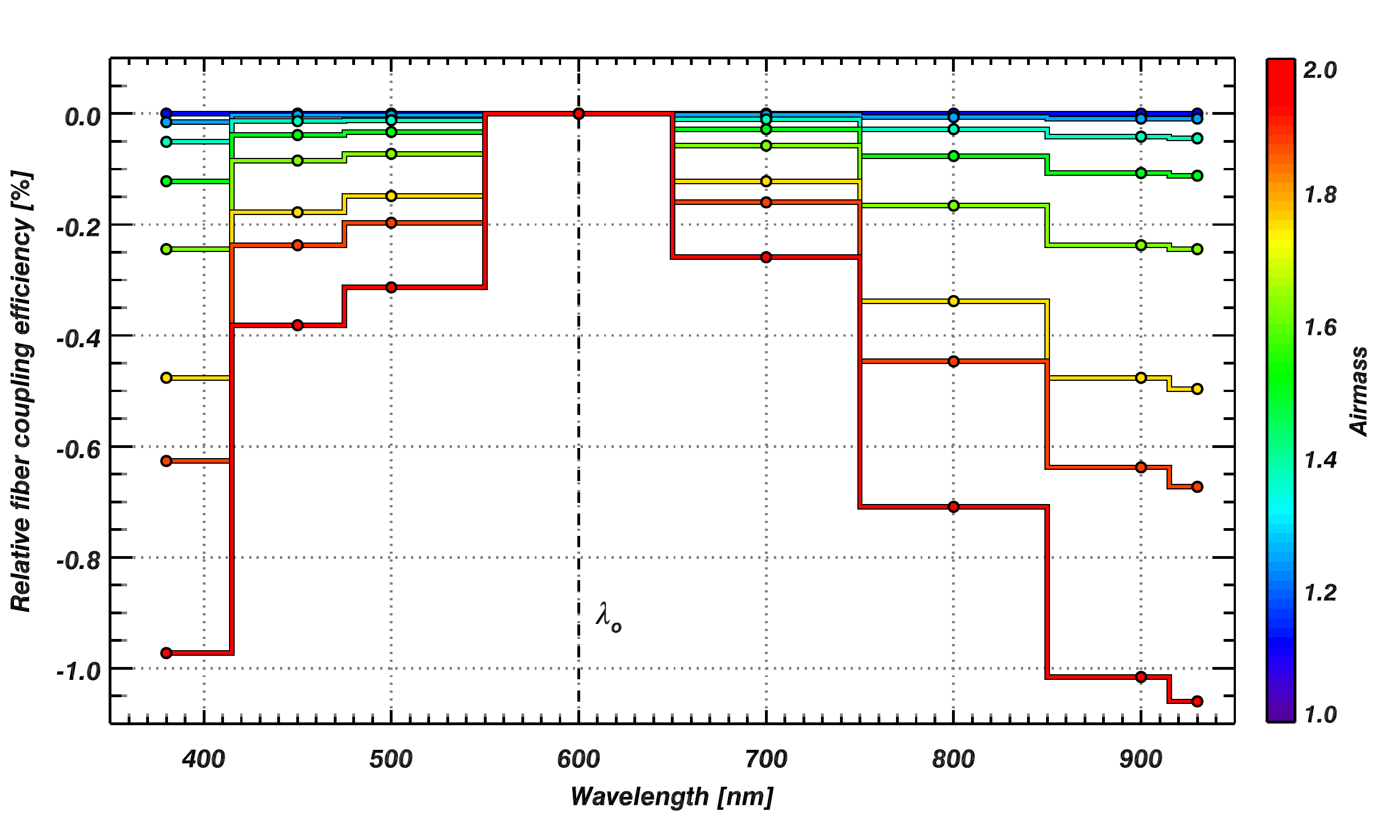}
\includegraphics[width=5.2in]{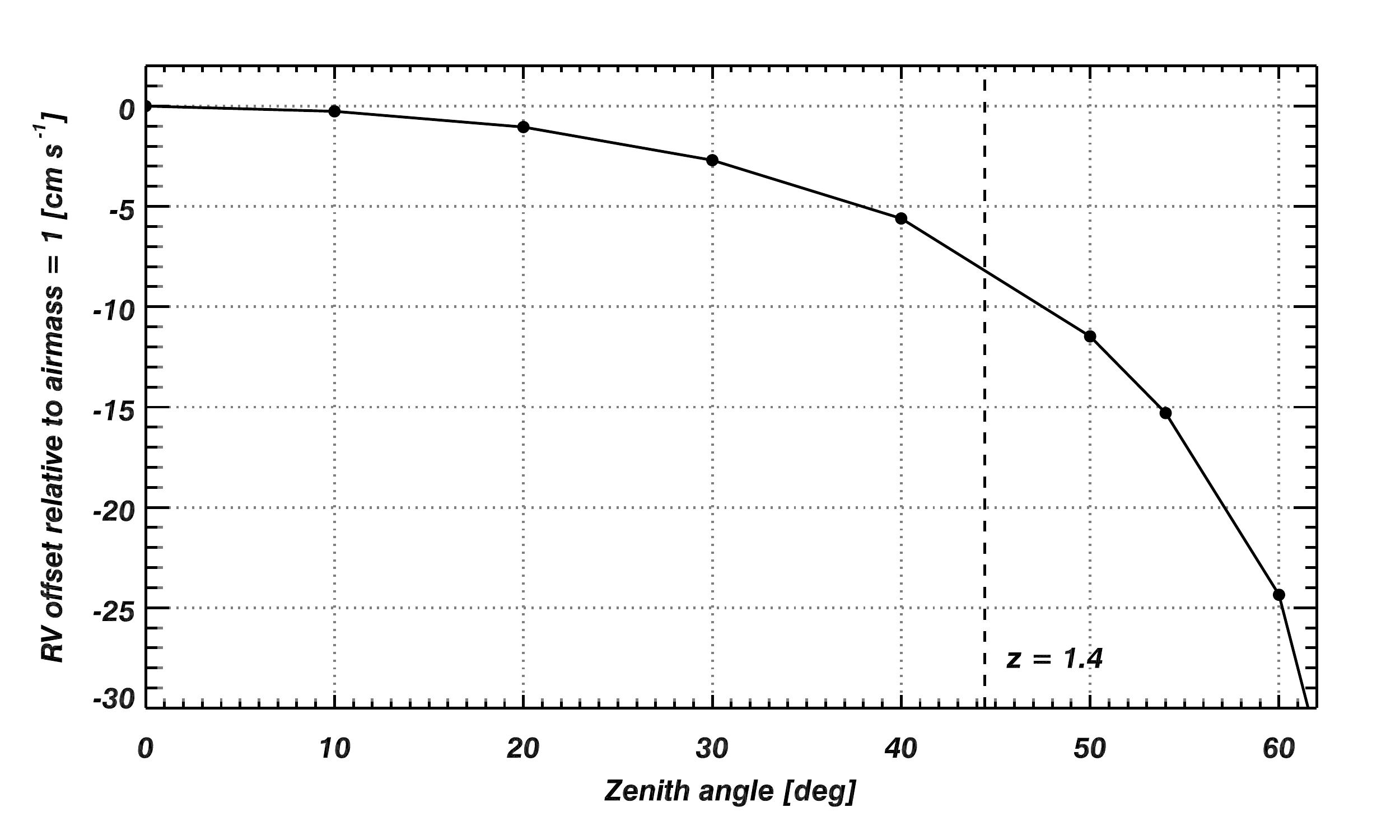}
\caption{Top: Theoretical chromatic fiber coupling efficiency as function for airmass for the notional NEID atmospheric dispersion corrector design. A central guiding wavelength of 600 nm is assumed, with eight wavelengths sampled across the full 380 -- 930 nm bandpass. The relative coupling efficiencies modulate the recorded stellar spectrum differently for different airmass values. Bottom: Calculated velocity offsets for stellar spectra weighted by the modulation curves shown above. This variable modulation of the stellar spectral energy distribution can introduce systematic RV errors if not properly corrected in the reduction software via repeatable continuum normalization.}
\label{fig:adc_efficiency}
\end{center}
\end{figure}

Variations in atmospheric dispersion will impact the measured velocities in two ways. Changing chromatic dispersion will result in: 1) variable chromatic coupling efficiency of light entering the fiber at the telescope focal plane (see Figure~\ref{fig:adc_efficiency} top), and 2) mimic guiding errors for wavelengths outside of the central guiding wavelength. The latter is particularly insidious, as it cannot be corrected in software post-facto and affects separate wavelengths in different ways.

To estimate the magnitude of the former effect on the achievable Doppler measurement precision, we modulate synthetic stellar spectra by the variable chromatic efficiencies expected for a high performance atmospheric dispersion corrector (ADC) at a variety of different airmasses. Effective velocity offsets due to the different chromatic weights are then calculated relative to a master template spectrum (see Figure~\ref{fig:adc_efficiency} bottom). Advanced analysis algorithms may have the ability to mitigate this effect through uniform and repeated flux weighting of measured spectral energy distributions \citep{Berdinas:2016, Halverson:2015b}, but we carry a conservative estimate of this error until it can be definitively shown to be calibratable through software at the c{\ms} level. We hold 6.5 c{\ms} error for this term for reasonable median airmass values ($\sim$1.4) based on the results shown in Figure~\ref{fig:adc_efficiency}.

For the chromatic image smearing component of the ADC error (analogous to chromatic scrambling error), a 0.1{\arcsec} peak-to-valley spread over the full NEID bandpass (consistent with the current notional ADC model up to airmasses of $\sim$2) leads to an effective 2.5 c{\ms} RV offset across the spectrum for our expected fiber scrambling gain. This term, combined in quadrature with the variable chromatic weighting term ($\sim$6.5 {c{\ms}}), leads to an estimated 6.9 c{\ms} overall error contribution from the ADC.

\subsubsection{Defocusing errors}
Variable (achromatic) telescope focus will directly affect the fiber coupling efficiency for the instrument. The defocused telescope image may also result in a net shift in the centroid illumination on the fiber face, though this is highly dependent on the expected focus depth and injection beam properties.  The net effect of these variations on achievable RV precision is difficult to estimate, though these effect can be mitigated with real-time focus control, high levels of image scrambling, and a precise exposure meter. We loosely hold 5 c{\ms} error to accommodate for defocusing effects, though further studies will be done once the initial performance of the NEID front-end system is evaluated.

\subsubsection{Windshake}

\label{subsubsec:windshake}
Transient vibrations of the telescope during high winds, also referred to as \lq{}windshake\rq{}, will manifest as large-scale guiding errors. This is particularly an issue for the WIYN telescope, as high winds are not uncommon during prolonged observation periods. Based on our expected scrambling performance for the NEID fiber train (see \S~\ref{sec:nf_scrambling}), an uncorrected 0.5{\arcsec} RMS guiding \lq{}jitter\rq{} on the fiber (indicative of the highest amplitude tracking jitter during heavy winds at the WIYN site) contributes roughly 8 c{\ms} RV error. This effect will be likely be reduced significantly with the high frequency tip-tilt system used in the NEID front end unit, and will also average out over long exposures to some extent. We conservatively carry 8 c{\ms} for this error, though it is important to note that this jitter manifests as a high frequency transient, and should not affect the majority of measurements.

\subsection{Calibration source}
This section outlines the error terms associate with the NEID wavelength calibration process. NEID will use a commercial Menlo Systems broadband optical frequency comb for precision Doppler calibration \citep{Molaro:2013, Probst:2014, Zou:2015}. The comb will provide a dense forest of stable emission features spanning the majority of the NEID bandpass (420 -- 900 nm, 20 GHz filtered mode spacing). Even with this fundamentally stable frequency reference, many sources of systematic error in the calibration process are still present at the c{\ms} level, and therefore must be included in our budget.

\subsubsection{Calibration process} 
\label{subsubsec:cal_snr} 
Signal-to-noise ratio variation in individual frequency comb calibration lines, combined with the finite pixel sampling of the spectrometer point spread function, results in measurable errors in line centroiding during the wavelength calibration process. These errors propagate to the wavelength solution with an algorithmic contribution of $<$5 c{\ms} based on simulations done for HPF optical frequency comb \citep{Terrien:2014}. We hold this value for our budget, though expect the improved NEID PSF sampling (5 pixels, rather than three for HPF) and exquisitely uniform spectral flattening inherent in the Menlo system ($<$3 dB across the bandpass) to reduce this contribution significantly.

\begin{figure}
\begin{center}
\includegraphics[width=4.8in]{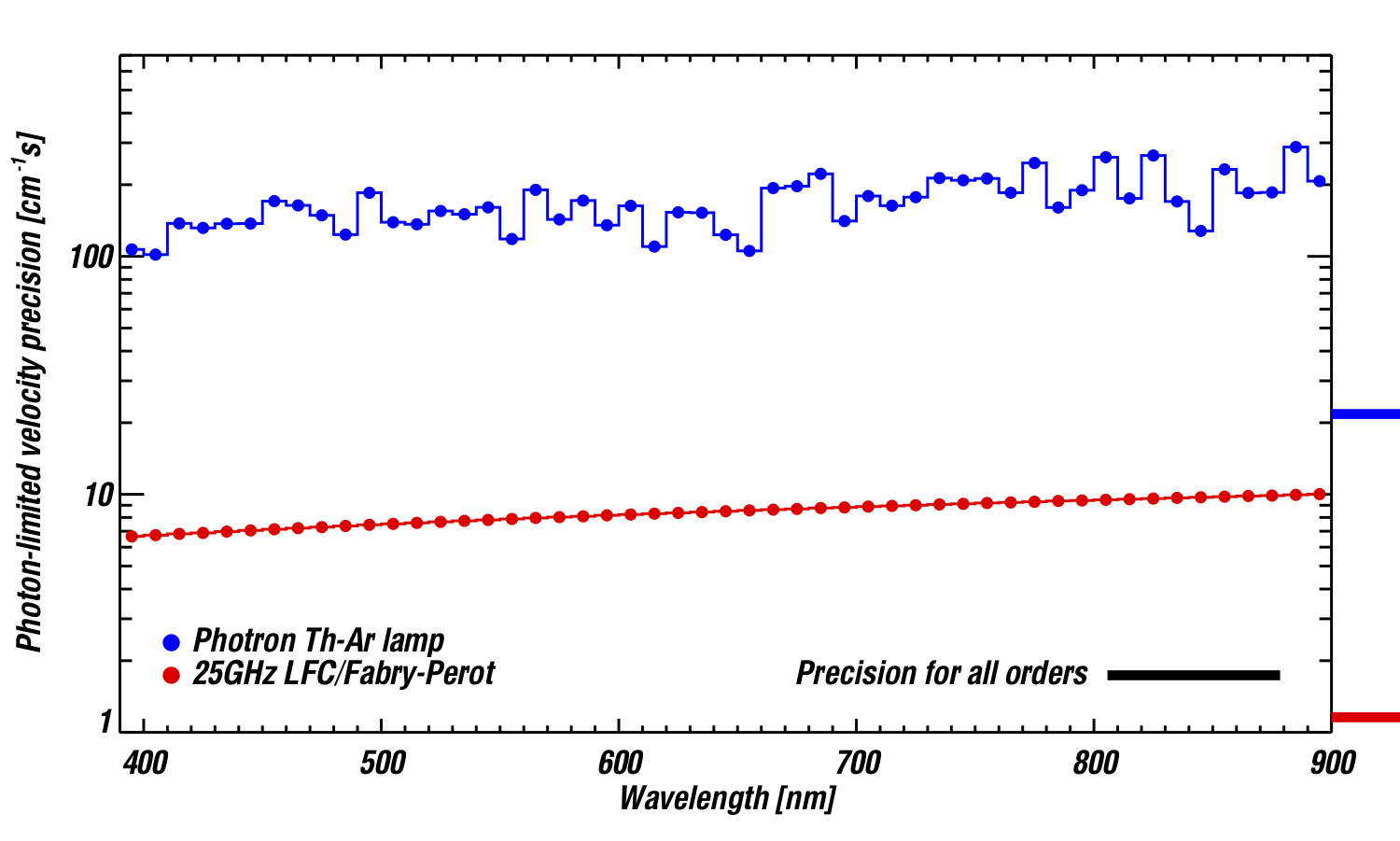}
\caption{Theoretical photon-noise limited RV precision of different calibration sources in the NEID bandpass. Simulations assume a SNR of 500 per extracted pixel, and an instrument resolution of 100,000. The photonic references (LFC/{\fp}) have significantly more rich calibration spectra, leading to over an order of magnitude improvement in calibrator precision over classical Th-Ar lamps.}
\label{fig:cal_photon_err}
\end{center}
\end{figure}

\subsubsection{Calibration accuracy} 
This term accounts for the absolute accuracy of the intrinsic calibration spectrum, which includes both the inherent calibration source stability and expected photon noise contribution. Long term stability of the Menlo Systems optical frequency comb is measured to be better than 2 c{\ms} (based on simultaneous comparisons between two separate combs, Menlo Systems, priv. comm.). We hold twice this value in our performance budget to maintain reserve.

We will use neutral density filters to both balance comb flux with stellar flux and to minimize photon noise to $<$4 c{\ms} (see Figure~\ref{fig:cal_photon_err} for example photon noise calculation). Flux matching is essential for maximizing dynamic range of possible observations and mitigating certain detector-related systematics.

\subsection{Software \& algorithms}
\label{subsubsec:software} 

Small algorithmic errors can affect the consistency of the reduction pipeline at these RV precisions. We hold 10 c{\ms} RV error for this term, which includes effects that are difficult to directly link to RV degradation, such as order tracing shifts, small scattered light changes, flat field uniformity, cosmic ray removal, choice of numerical template mask lines used for cross-correlation, and the measurement of the cross-correlation function center. This estimate is based on extensive algorithmic modeling done for HPF using a custom-built software simulator \citep{Terrien:2014}, which included the state-of-the art extraction algorithms we plan to adapt directly to NEID. 

\subsection{Atmospheric sources}
Beyond simply restricting wavelength regions available for precision Doppler measurements, atmospheric contamination plays a key role in setting the achievable measurement floor. Here we summarize our efforts to quantify the major contributions of atmospheric contamination.

\subsubsection{Micro-telluric contamination}
\label{subsubsec:micro_telluric}

\begin{figure}
\begin{center}
\includegraphics[width=4.8in]{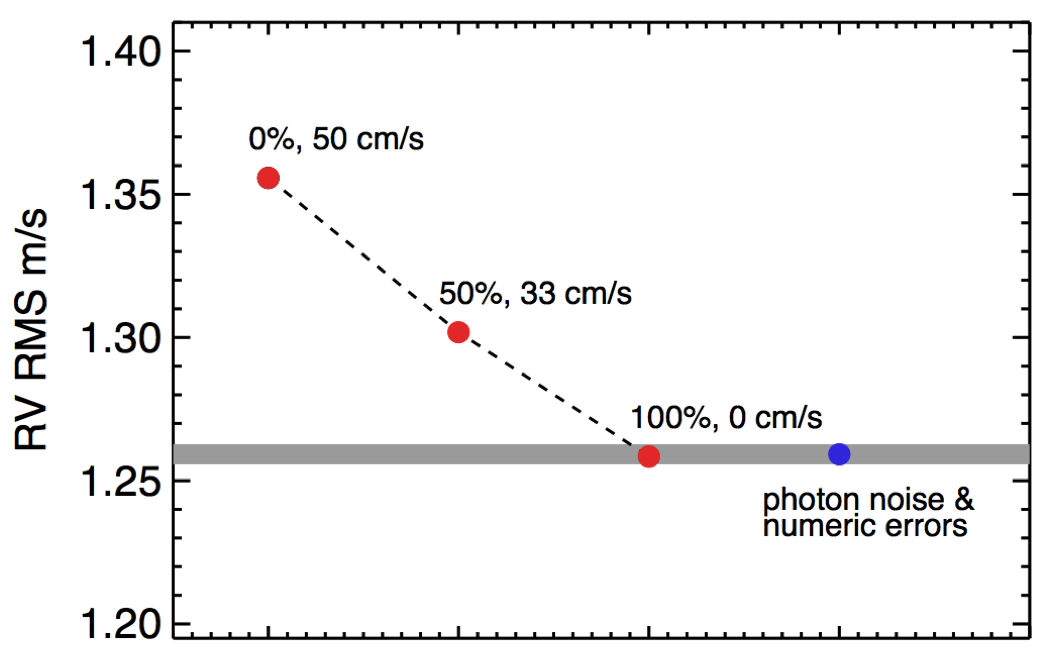}
\caption{Expected measurement precision improvement as function of micro-telluric subtraction level for simulated Keck/HIRES velocity measurements of an RV standard star. Micro-telluric features are modeled at varying levels of correction (between 0 and 100 \%). Current telluric modeling and removal capabilities enable subtraction at the $<$10\% level. }
\label{fig:microtelluric_ex}
\end{center}
\end{figure}

Small amplitude \lq{}micro-telluric\rq{} lines are atmospheric absorption features that are unresolved at the $<$1\% level. These features are particularly insidious for RV work, and populate virtually the entire NEID bandpass to some extent. These features are generally undetected in recorded spectra, though will introduce systematic RV measurement error by imposing an aliasing signal at harmonics of the sidereal year. \cite{Artigau:2014b} and \cite{Cunha:2014} used forward-modeling techniques and synthetic telluric atmospheres to reduce the micro-telluric contamination in HARPS data. Based on their work, and the detailed simulations done within our NEID collaboration on modeling this effect for synthetic Keck/HIRES data (see Figure~\ref{fig:microtelluric_ex}), we estimate the uncorrectable micro-telluric error contribution in NEID (due to fast winds, variable water column, and modeling uncertainties) to be 10 c{\ms}.

\subsubsection{Sky fiber subtraction}
\label{subsubsec:sky} 
\begin{figure}
\begin{center}
\includegraphics[width=5.2in]{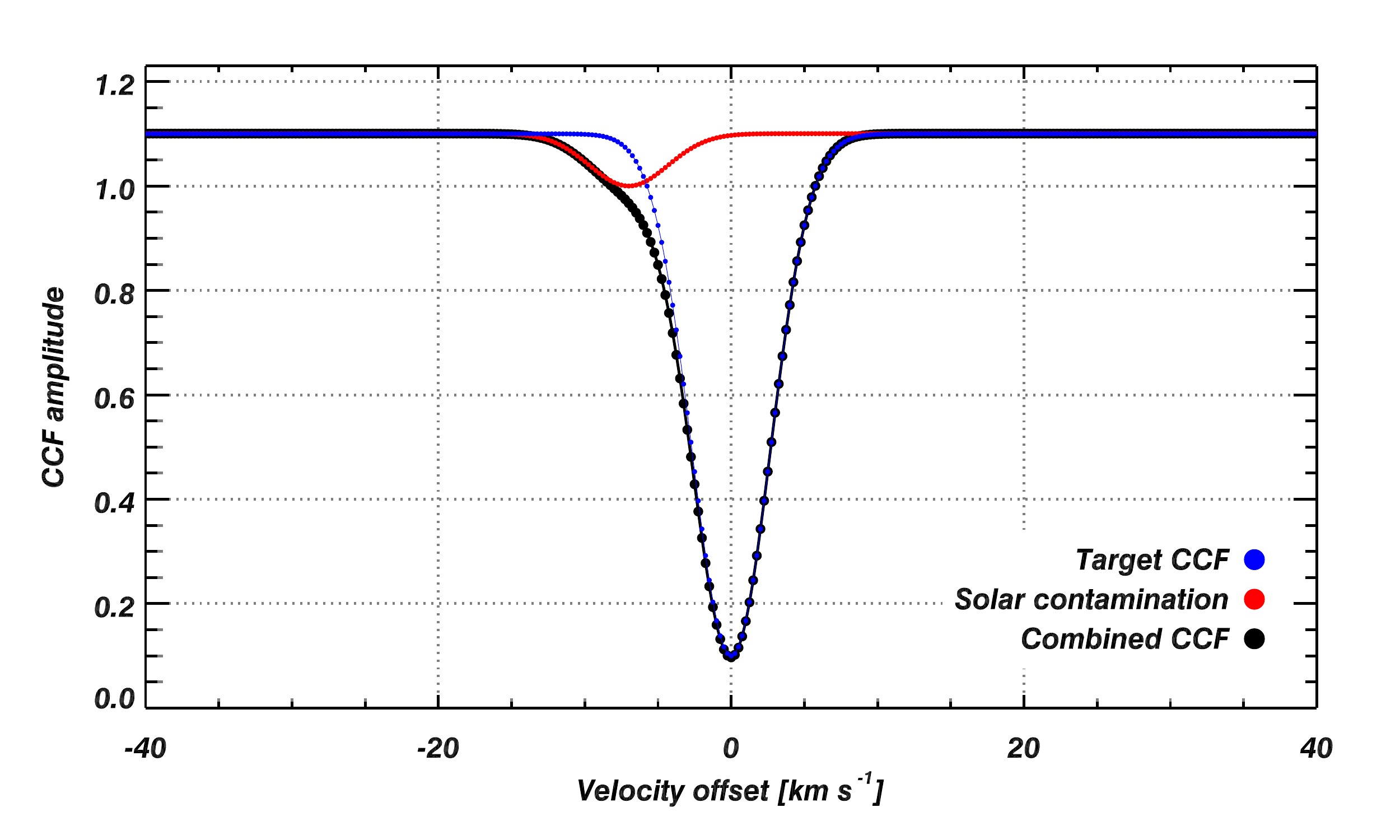}
\caption{Exaggerated example of Solar contamination affecting RV measurement. A velocity-shifted solar spectrum will be imprinted on any measured stellar spectrum, leading to an asymmetry in the velocity cross-correlation function (CCF). Even during observation periods with minimal moonlight contamination and low-levels of atmospheric scattering, solar contamination can still introduce systematic measurement errors in the CCF fitting process at the 10's of c{\ms} level. The NEID sky fiber, with accompanying telluric modeling software, will allow for solar contamination correction at the $\sim$1 \% level.}
\label{fig:solar_contamination}
\end{center}
\end{figure}

Solar contamination, due to moonlight and atmospheric scattering, can cause additional systematic RV noise. The magnitude of this effect depends on median site sky brightness, target-moon separation on the sky, lunar phase during time of observation, ecliptic latitude, zenith angle, and phase of the solar cycle (e.g., \cite{Krisciunas:1997}). To quantitatively asses the impact of solar contamination on the measured RVs, we simulate the output cross-correlation function of a recorded stellar spectrum combined with the solar spectrum at realistically varying velocity separations and amplitudes (see Figure~\ref{fig:solar_contamination} for an exaggerated example of CCF contamination from Solar features). 

The solar spectrum is scaled to the Kitt Peak dark time sky brightness (V=20.9 mag/arcsec$^2$, airmass=2, \cite{Massey:2000}), with additional moon contamination of 1.8 mag/arcsec$^2$ (characteristic of \lq{}grey\rq{} time at the WIYN site). Left uncorrected, this contamination directly affects the measured RVs, with an RMS RV error of $\sim$2 {\ms} for a V=12 mag star. Under bright conditions, ($\sim$17 mag/arcsec$^2$ median sky brightness), the error worsens drastically to $\sim$13 {\ms} if not subtracted properly. For a V=16 mag star, the precision degrades further to 90 -- 150 {\ms}. Our simulations show that the inclusion of a sky fiber, which gives a direct probe of the brightness of the contaminating solar spectrum, in our NEID design is critical to meeting the instrument's bold precision goal. Sky subtraction or modeling resulting in residuals of 1\% or better reduces the RV contamination to 3 c{\ms}; 5\% residuals correspond to 10 c{\ms} (assuming V=12 target). We hold 10 c{\ms} error for this term to maintain margin until further constraints can be placed.

\subsection{Barycentric correction errors}
\label{subsubsec:bc}
Measuring accurate stellar RVs requires accounting for the barycentric motion of the telescope due to the Earth's motion through the Solar System. Sophisticated algorithms and ephemerides, combined with the known observatory positions and precise astrometric measurements, can determine the barycentric correction (BC) to $<$1 c{\ms} \citep{Wright:2014}, given error-free inputs. However, errors in the effective time of observation of only 2 seconds can result in a BC error of 1 c{\ms}. Flux variations caused by guiding errors, seeing variations, and changes to the atmospheric transmission (e.g. due to variable cloud cover, changes in airmass, etc.) are all potential sources of timing error at these levels. Deriving an accurate BC requires that we monitor the flux entering the spectrometer as a function of time, and compute a \lq{}flux-weighted\rq{} barycentric correction. For NEID we employ a chromatic exposure meter to reduce the worst case error contribution from this effect to $\sim$0.8 c{\ms} for G stars. Measuring chromatic flux time series will be crucial for minimizing this error term, as variable guiding, ADC performance, and atmospheric conditions will affect different wavelengths at variable levels.

\subsection{Cross correlation function masks} 
\begin{figure}
\begin{center}
\includegraphics[width=4.8in]{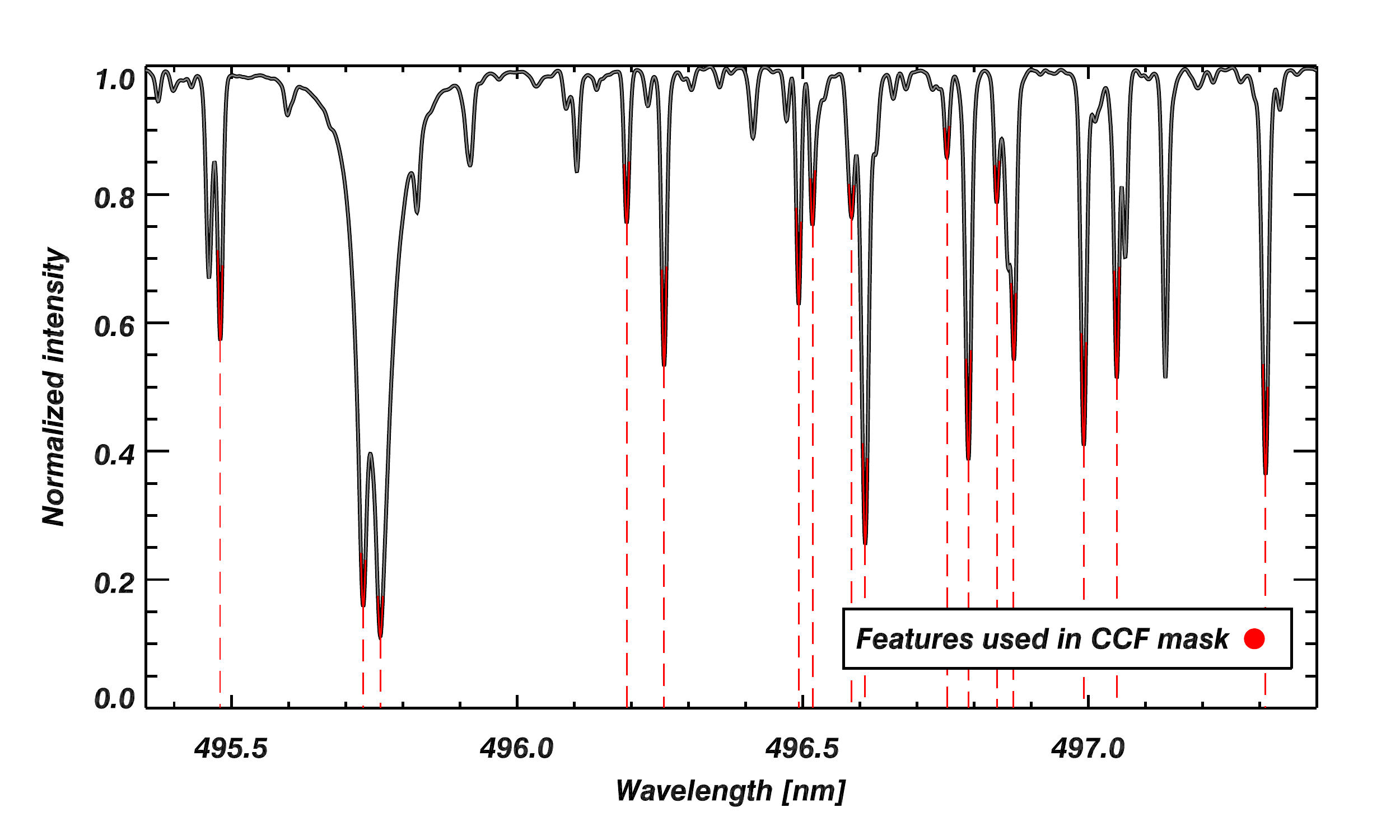}
\caption{High resolution sample of Solar spectrum showing individual features used in cross-correlation mask (features marked with dashed lines). Many features in the spectrum are not used to derive stellar velocities, though the exclusion of these features has little impact on the achievable measurement precision.}
\label{fig:mask_photon}
\end{center}
\end{figure}

For NEID, RVs will be measured using a CCF mask technique, which effectively masks out regions with little or no useful RV information (e.g., continuum). This technique naturally excludes many regions that would otherwise contribute minimal, albeit non-negligible, information content (see Figure~\ref{fig:mask_photon} as an example region with an overlaid spectral mask). To quantify the effect of using a wavelength mask, rather than using the entire recorded spectrum, we compared the photon noise calculations for synthetic stellar spectra with and without custom spectral CCF masks. We tested three different spectral masks in the HARPS region (370 -- 680 nm), and found that inclusion of the mask degraded the theoretical photon-limited RV precision by $\sim$10\%, consistent across several spectral types in this wavelength range. This constant scaling informs our stellar photon-limited RV precision calculations which, while explicitly not included in the instrumental error budget presented here, does affect the fundamental floor for the achievable on-sky measurement precision.

\section{Conclusions}
We have developed a detailed, bottom-up Doppler radial velocity performance budget for the Extreme Precision Doppler Spectrometer concept NEID. This analysis includes many of the error terms and methodologies introduced in \cite{Podgorski:2014}, though we include many new sources of error in our Doppler budget. This budget contains a suite of contributions from a variety of instrumental error sources. While primarily focused on the performance of the NEID spectrometer, this systems engineering approach of performance budgeting could easily be applied to a number of future RV instruments to derive estimated performance. This type of analysis and systems engineering is crucial for illuminating a path towards 10 c{\ms} precision, which has long been held as the ultimate Doppler precision goal in the quest for discovering Earth-size planets orbiting Sun-like stars.

\acknowledgments{This work was partially supported by funding from the Center for Exoplanets and Habitable Worlds. The Center for Exoplanets and Habitable Worlds is supported by the Pennsylvania State University, the Eberly College of Science, and the Pennsylvania Space Grant Consortium. HPF development is supported by NSF grants AST 1006676, AST 1126413, and AST 1310885. NEID development is supported by JPL subcontracts 1531858 and 1547612. SPH acknowledges support from the Penn State Bunton-Waller and Braddock/Roberts fellowship programs, as well as the Sigma Xi Grant-in-Aid program. This work was performed in part under contract with the California Institute of Technology (Caltech)/Jet Propulsion Laboratory (JPL) funded by NASA through the Sagan Fellowship Program executed by the NASA Exoplanet Science Institute.}

\bibliography{/Users/kicker/Desktop/bib_file/bib_master}
\bibliographystyle{apj} 

\end{document}